\documentclass[11pt]{article}
\usepackage{graphicx}
\usepackage{amsmath}
\usepackage[usenames]{color}
\def\mi{\bigskip\noindent}
\def\deg{{^\circ}}
\newcommand{\Ro}{\text{Ro}}
\newcommand{\E}{\text{E}}
\newcommand{\Rm}{\text{Rm}}
\newcommand{\Pm}{\text{Pm}}
\newcommand{\Ha}{\text{Ha}}
\renewcommand{\Re}{\text{Re}}
\newcommand{\N}{\text{N}}
\begin{document}
{\center

\bigskip
{\bf Experimental study of super-rotation in a magnetostrophic spherical Couette flow}

\mi H-C. Nataf \footnote{email: Henri-Claude.Nataf@ujf-grenoble.fr}, T. Alboussi\`ere, D. Brito, P. Cardin, N. Gagni\`ere, D. Jault, J-P. Masson and D. Schmitt  

\mi Geodynamo team, LGIT-UMR5559-CNRS-UJF, Grenoble, France

\bigskip
submitted to {\it Geophysical and Astrophysical Fluid Dynamics}\\
\today

}

\bigskip{\bf Abstract.} We report measurements of electric potentials at the surface of a spherical container of liquid sodium in which a magnetized inner core is differentially rotating. The azimuthal angular velocities inferred from these potentials reveal a strong super-rotation of the liquid sodium in the equatorial region, for small differential rotation. Super-rotation was observed in numerical simulations by Dormy et al. \cite{dor98}.
We find that the latitudinal variation of the electric potentials in our experiments differs markedly from the predictions of a similar numerical model, suggesting that some of the assumptions used in the model -- steadiness, equatorial symmetry, and linear treatment for the evolution of both the magnetic and velocity fields -- are violated in the experiments. In addition, radial velocity measurements, using ultrasonic Doppler velocimetry, provide evidence of oscillatory motion near the outer sphere at low latitude: it is viewed as the signature of an instability of the super-rotating region. 


\section{Introduction}

In the recent years a revived interest has been shown for the rotating spherical Couette flow: the flow between two concentric spheres that rotate around a common axis $z$ with different angular velocities $\Omega$ (outer sphere) and $\Omega+ \Delta\Omega$ (inner sphere). Among the reasons for this renewed interest are the possibility of super-rotation when a magnetic field is added, and the prospect of observing dynamo action with such a flow.

The first studies \cite{pro56} dealt with the axisymmetric flow in the limit $\Ro = \Delta\Omega / \Omega \rightarrow     0$ and $\E = \nu / \Omega a^2 \rightarrow 0$, where $\Ro$ and $\E$ are the Rossby and Ekman numbers, respectively, $\nu$ is the kinematic viscosity of the fluid, and $a$ the radius of the outer sphere. Proudman \cite{pro56} discovered that, under the action of the Coriolis force, a vertical shear layer sets in to accommodate the transition between the fluid at rest (in the frame rotating at $\Omega$) outside the vertical cylinder tangent to the inner sphere, and the fluid rotating at an angular velocity ranging from  $\Omega+ \Delta\Omega$  to  $\Omega+ \Delta\Omega/2$ inside the tangent cylinder. Stewartson \cite{ste66} unraveled the inner structure of this shear layer for small yet finite values of the Ekman number, showing that an ageostrophic layer of width $a\E^{1/3}$ is nested between two geostrophic layers of width $a\E^{2/7}$ (inwards) and $a\E^{1/4}$ (outwards). Years later, the validity domain of these analytical results was established in numerical simulations \cite{hol94,dor98}.

The Stewartson shear layer is expected to become unstable to non-axisymmetric perturbations when the Rossby number is increased above a critical value $\Ro_{c}$. Experimental evidence was found by Hide and Titman \cite{hid67}. These instabilities were also studied in 3D-numerical simulations by Hollerbach \cite{hol03} who emphasized the difference between the $\Ro>0$ and $\Ro<0$ cases. When the Ekman number is small enough then $\Ro_c \ll 1$, and experiments demonstrate that the instabilities are $z$-independent \cite{nii84,sch05a}. Schaeffer and Cardin \cite{sch05a} thus conducted a quasi-geostrophic analysis, and showed that the instabilities are akin to Rossby waves, thereby explaining the dependence on the sign of $\Ro$. They found that the onset scales as: $\Ro_c \propto \beta \E^{1/2}$, where $\beta$ is associated to the slope of the sphere at the latitude of the Stewartson layer. Above the threshold, a strong anisotropic and inhomogeneous Rossby wave turbulence develops, which exhibits a rather steep $m^{-5}$ spectrum of kinetic energy in the azimuthal direction \cite{sch05b} ($m$ is the azimuthal number). It means that a direct cascade of energy is prohibited by the geostrophic constraint, leading to a very low kinetic energy of the fluid flow at small scales.

Once the instabilities are strong enough and when the fluid is an electric conductor, the instabilities can produce a magnetic field through dynamo action \cite{sch06}. This happens when the magnetic Reynolds number $\Rm = U a/\eta$  is larger than some critical value $\Rm_c$ (where $U$ is the typical flow velocity of the fluid and $\eta$ its magnetic diffusivity). Within the quasi-geostrophic approach, $\Rm_c$ is found to be in the range of 3000. However, full 3D numerical simulations yield much smaller values around 100 \cite{wic06}, but they are limited to larger values of both the Ekman number and the magnetic Prandtl number $\Pm=\nu/\eta$. These results validate the project of building an experimental dynamo based on the rotating spherical Couette flow \cite{car02}. After the success of the liquid sodium dynamos in Riga \cite{gai01} and Karlsruhe \cite{sti01}, such a flow would get us one step closer to situations relevant to planetary cores.

\begin{figure}
  \centerline{\includegraphics[width=5cm]{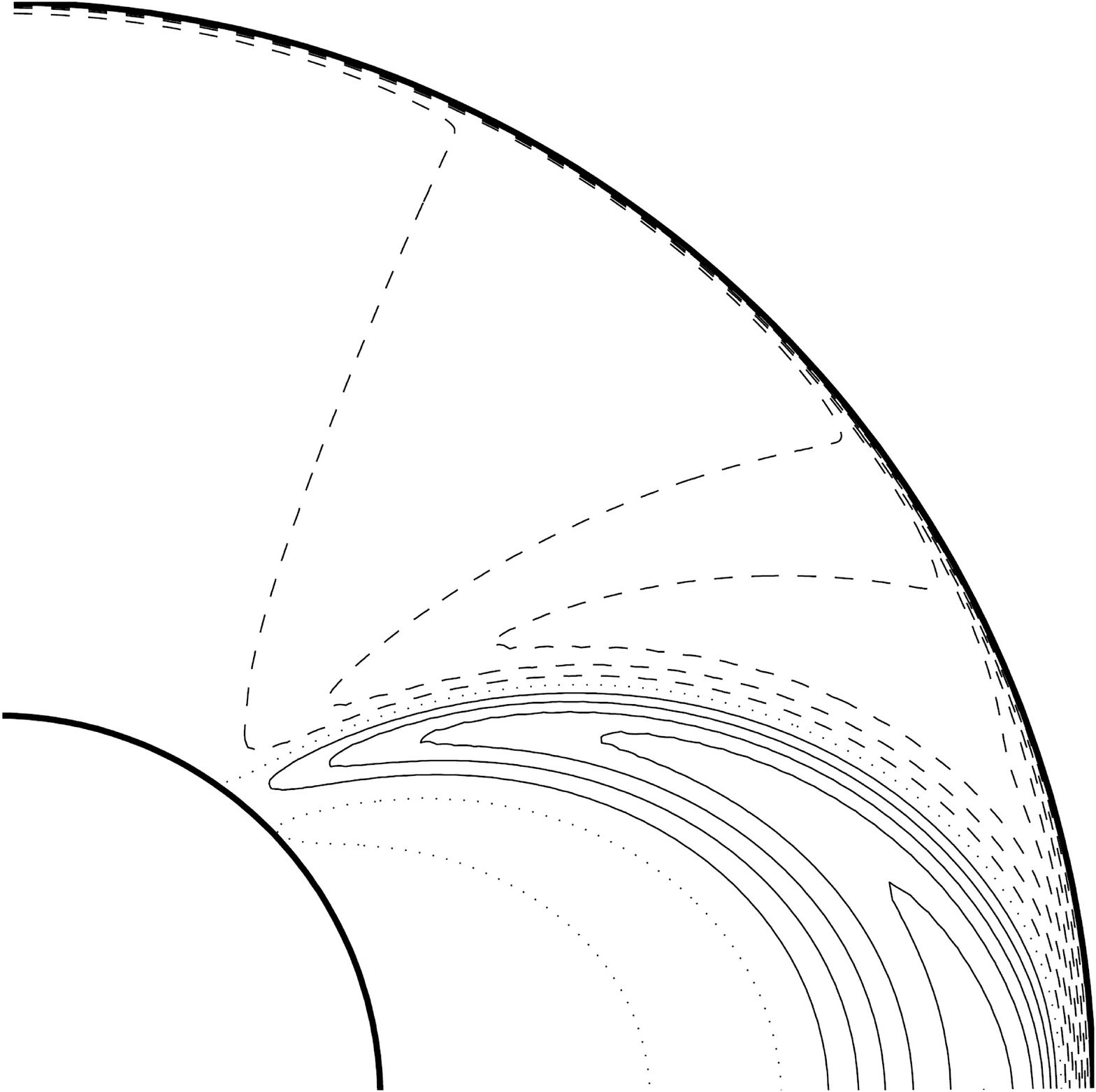} \includegraphics[width=5cm]{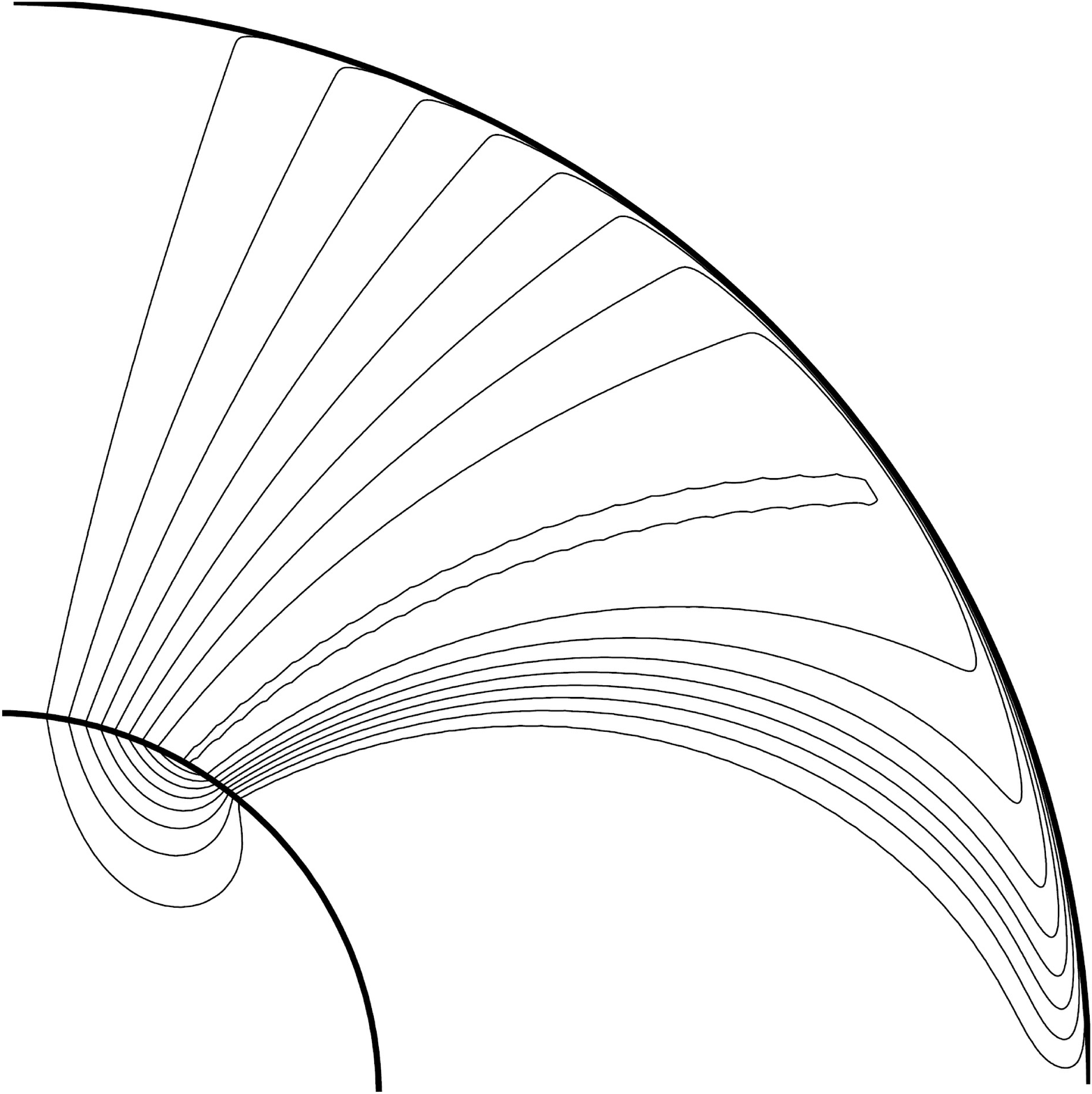} \includegraphics[width=5cm]{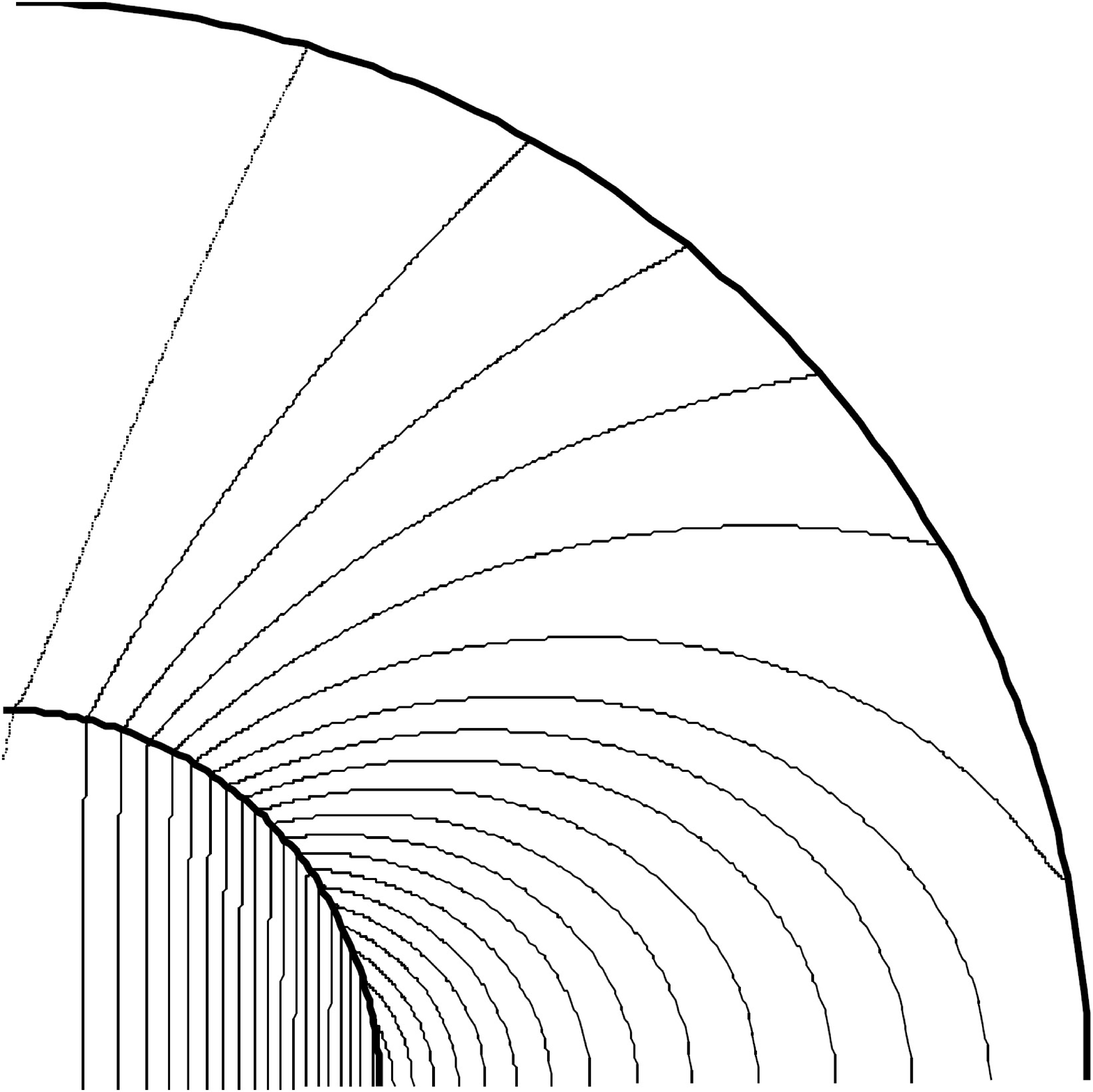}}
    \begin{picture}(0,0) (0,0)
    \put(0,25){(a)}
    \put(150,25){(b)}
    \put(290,25){(c)}
  \end{picture}
    \caption{Super-rotation in axisymmetric numerical simulations (outer sphere at rest). The numerical method is described in \cite{dor02} and has been extended to take into account the finite electrical conductivities of the outer shell and inner sphere of the $DTS$ experiment. The strength of the magnetic field (Hartmann number) is also the same as in $DTS$ (see Table \ref{tab:dimen}). Isovalue contours in a meridional plane of (a) the normalized angular velocity $U_{\varphi}/s \Delta \Omega$ (maximun 1.53, contour level 0.1, dashed line for value lower than 1, dotted line for value 1 and continous line for value larger than 1); (b) the dimensionless stream function of meridional electric currents (maximun 0.93, contour level 0.1) -- the electric currents are clockwise; (c) the normalized electric potential $V/a^2 \Delta \Omega B_o$ in the frame of the outer sphere (maximum 600, contour interval 30).}
        \label{fig:num}
\end{figure}

However, the presence of a magnetic field $\vec{B}$ will have a strong influence on the flow itself. This is well illustrated by the numerical studies of Hollerbach \cite{hol94} and Dormy et al. \cite{dor98}, who investigate rotating spherical Couette flow with an imposed magnetic field. Both studies only deal with the axisymmetric solution obtained in the limit $\Ro\rightarrow 0$ ($(\vec{u}\cdot\vec{\nabla})\vec{u}$ terms neglected in the Navier-Stokes equation) and $\Rm\rightarrow 0$ (effects of the induced magnetic field are neglected). One effect of the magnetic field is to thicken the Stewartson layer. For large enough magnetic fields, the quasi-geostrophic character of the flow is broken. More intriguing, Dormy et al. \cite{dor98} discovered that, when the imposed magnetic field is an axial dipole, the fluid can rotate at an angular velocity larger than that of the inner sphere, in particular when the outer sphere is at rest. This phenomenon of super-rotation is illustrated in Figure \ref{fig:num}, which shows the dimensionless angular velocities ($U_{\varphi}/s \Delta \Omega$, where $U_{\varphi}$ is the azimuthal velocity and $s$ the cylindrical radius), meridional electric currents and electric potentials in a meridional plane. A wide crescent of super-rotation is clearly visible in Figure \ref{fig:num}a. This phenomenon may be explained using the circulation of electric currents (Figure \ref{fig:num}b) : the differential rotation generates a viscous shear at the outer boundary and electric currents are induced in the shear layer (named Hartmann layer) so that Lorentz forces balance the viscous forces. The electric currents of the Hartmann layers loop into the bulk of the fluid. They tend to follow the dipolar magnetic field lines but this is not possible in the equatorial region where they have to cross the magnetic field lines. In this region, they generate Lorentz forces that accelerate to a super rotating angular velocity a ring of fluid around the magnetic field line tangent to the outer sphere. An asymptotic analysis of this peculiar phenomenon has been carried out by Dormy et al. \cite{dor02} for $\Omega =0$ (also see \cite{kle97} and \cite{sta98}). It indicates that the width of the detached shear layer varies as $\Ha^{-1/2}$, while the thin layers on the rigid surfaces are classical Hartmann layers \cite{mor90} whose thickness varies as $\Ha^{-1}$ (the square of the Hartmann number $\Ha$ characterizes the ratio of electromagnetic forces to viscous forces, and is defined in table \ref{tab:dimen}).

Hollerbach \cite{hol00,hol01} showed that the resulting flow and the phenomenon of super-rotation strongly depend upon both the geometry of the applied magnetic field (axial versus dipolar for example) and the electric boundary conditions on the inner and outer spheres. In particular, when both boundaries are perfectly conductive and for an axial homogeneous magnetic field, Hollerbach \cite{hol01} and Hollerbach and Skinner \cite{hol01b} observe that the fluid rotates in the direction opposite to that of the driving inner sphere, and that the amplitude of this counter-rotation increases with the strength of the applied magnetic field. Buehler \cite{bue04} provided a simple explanation for this phenomenon: he showed that the flow rate in the counter-rotating jet is independent of the Hartmann number (provided it is large enough), and since the width of the jet decreases as $\Ha^{-1/2}$, its velocity must increase as $\Ha^{1/2}$. 

In contrast, with a dipolar magnetic field and an insulating outer sphere, the super-rotation is positive and remains below $1.4$ for $\Omega = 0$ \cite{dor98}. Hollerbach and Skinner \cite{hol01b} also studied the linear stability of the axisymmetric jets for an applied axial magnetic field. Whatever the electrical boundary conditions, they predicted the onset of instability for Reynolds numbers of a few thousands and their non-linear calculations reveal a few drifting vortices aligned with the imposed magnetic field.

When global rotation is present ($\Omega \neq 0$), a competition takes place between the Lorentz and Coriolis forces. As the Elsasser number $\Lambda = \sigma B^2/\rho \Omega$   decreases ($\sigma$ the electrical conductivity and  $\rho$ the density of the fluid), the ring of super-rotation stretches in the vertical direction and gets closer to the tangent cylinder, while super-rotation decreases \cite{dor98}. The magnetostrophic regime, in which both forces are comparable is of particular interest, since it is the regime expected for planetary dynamos (e.g. \cite{gub87}).
The above analyses always consider very low magnetic Reynolds numbers where the non-linear action of the induced magnetic field is neglected. In an attempt to retain non-linear contributions in the Lorentz force ($\Rm \neq 0$), while neglecting those in the advection term ($\Ro\rightarrow 0$), Cupal et al. \cite{cup03} found a substantial increase of super-rotation up to 3.5 in numerical simulations.

Little is known when all non-linear effects are present. Is super-rotation still present? How is the angular velocity distributed? Which symmetries are preserved? What are the characteristics of the meridional circulation? 

In this paper, we present the first experimental measurements of velocities in a magnetostrophic spherical Couette flow in presence of a strong dipolar magnetic field. Azimuthal velocities in the liquid sodium flow are obtained by measuring the electric potential at the outer surface. Radial profiles of the radial velocity are measured with an ultrasonic Doppler velocimeter. A wide range of parameters is explored, with magnetic Reynolds number $\Rm$ up to 25.
Experiments in a similar geometry have been conducted with a weak axial magnetic field and without global rotation \cite{sis04,spe06}. Turbulent liquid sodium flows in cylinders have also been investigated \cite{vks02}.

The $DTS$ set-up and measurement techniques are described in section 2. The relevant dimensionless numbers are listed in table \ref{tab:dimen}. Section 3 is devoted to our experimental results, which are discussed in section 4.

\section{Experimental set-up}

\begin{figure}
  \centerline{ \includegraphics[width=13cm]{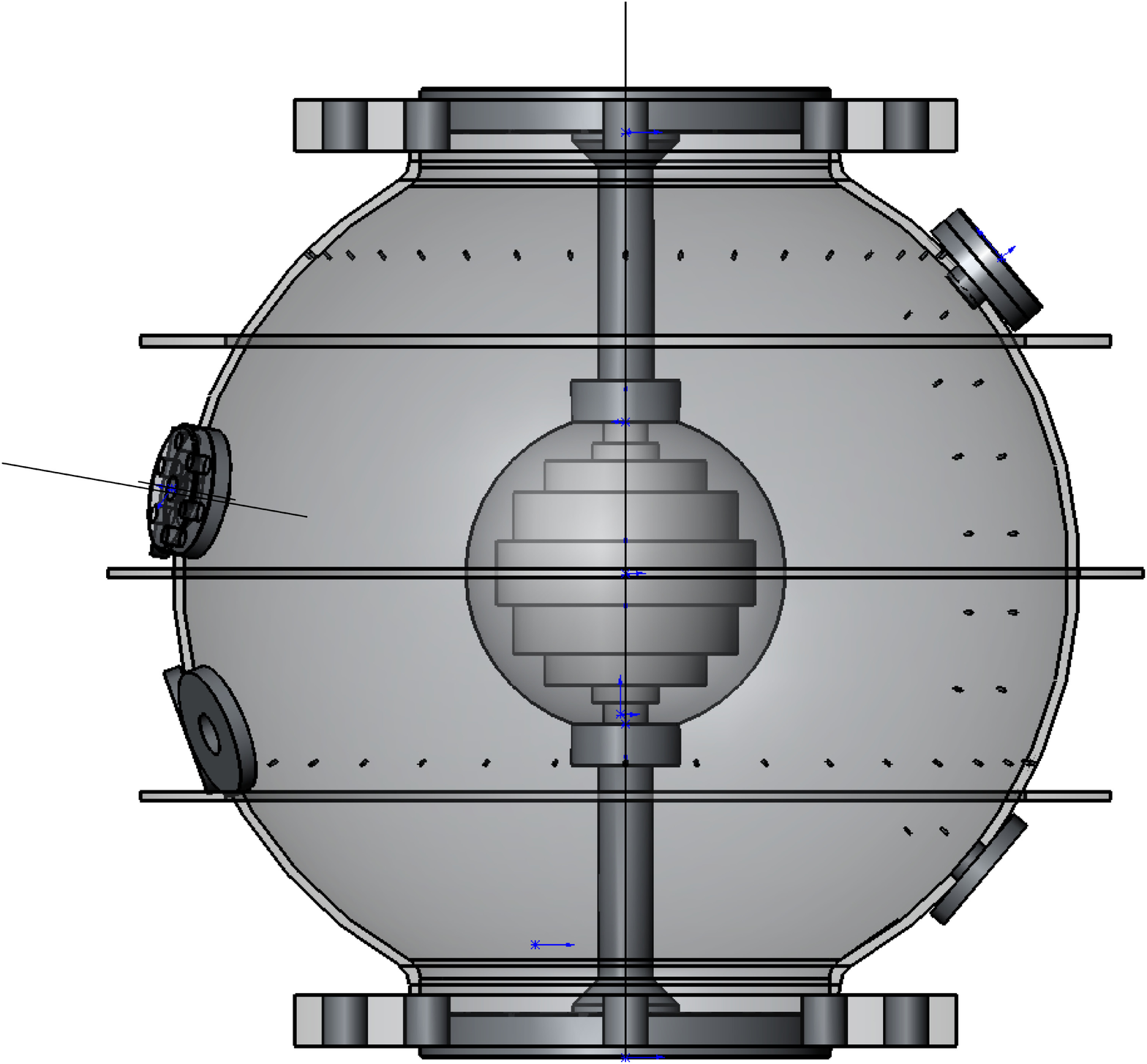}}
  \vskip -12 cm
  \includegraphics[width=4cm]{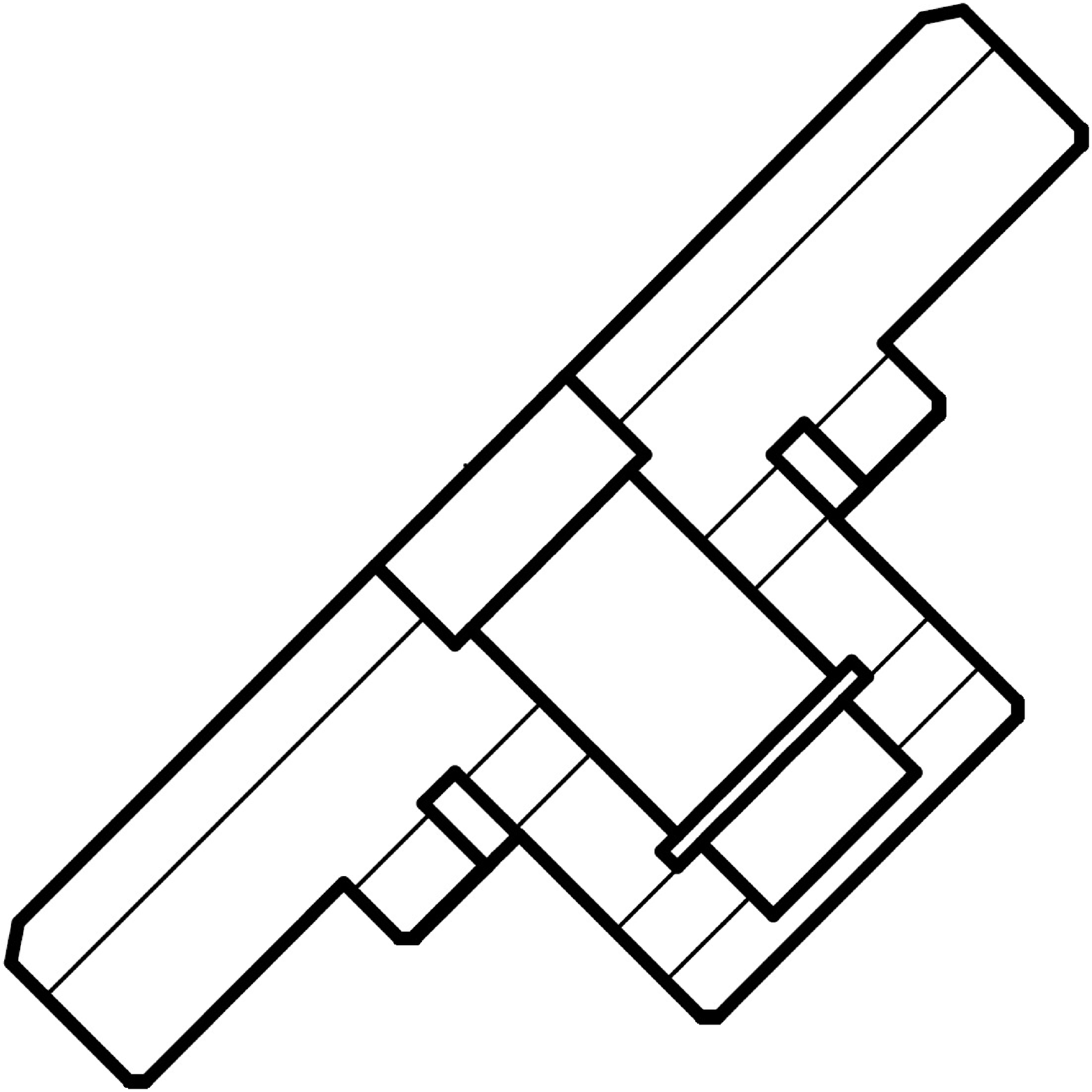}
  \hskip 9.5 cm
  \includegraphics[width=2cm]{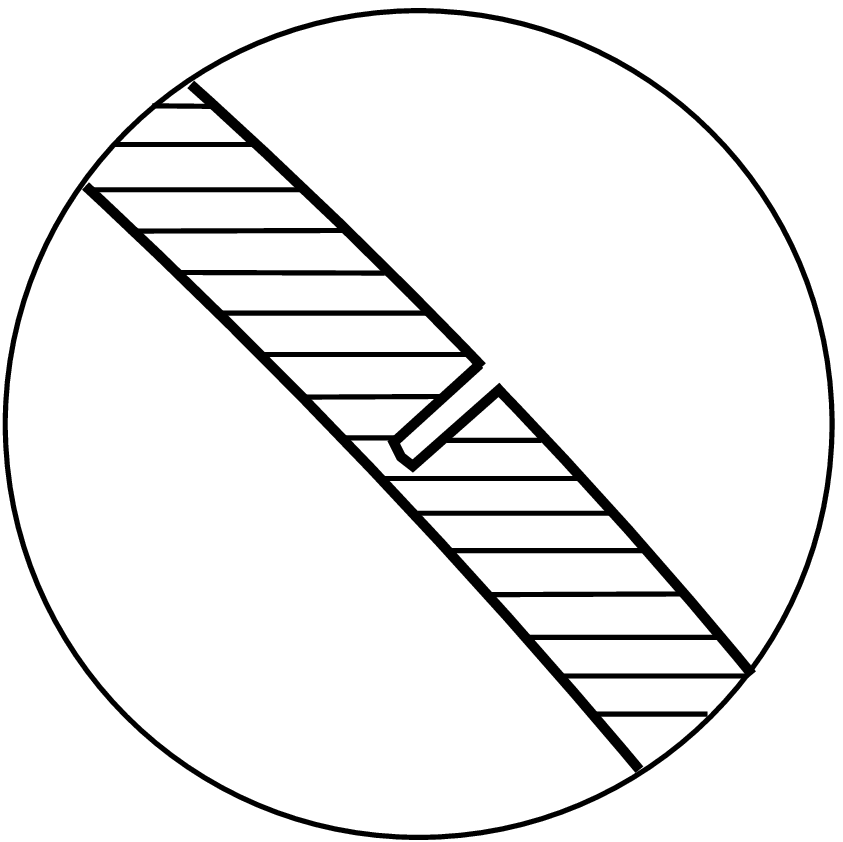}
  \vskip 8 cm
  \begin{picture}(0,0) (0,0)
    \put(140,200){sodium}
    \put(215,143){magnet}
  \end{picture}
    \caption{Sketch of the core of the $DTS$ experiment. The 7 cm-radius inner sphere is made of copper and contains 5 layers of magnets.  The inner radius of the stainless steel outer shell is 21 cm and its thickness is 5 mm. Tiny holes are drilled along several meridians and parallels to receive copper electrodes. They are 1 mm-diameter and 4 mm-deep blind holes as shown in the blown-up cross-section on the right. The large holes at latitudes $-20 \deg$, $10 \deg$ and $\pm 40 \deg$ receive interchangeable assemblies; a cross-section through one of them is shown on the left. An ultrasonic transducer can be mounted inside it.}
        \label{fig:sketch}
\end{figure}

A schematic drawing of the central part of the $DTS$ experiment is shown in Figure \ref{fig:sketch}. Forty litres of liquid sodium are contained between a 7 cm-radius inner sphere and a 21 cm-radius outer spherical surface. The 5 mm-thick outer spherical shell is made of stainless steel, which is about 8 times less electrically conductive than sodium at 130$\deg$C, the typical experimental temperature. The relevant properties of liquid sodium at 130$\deg$C are given in Table \ref{tab:properties}. The inner sphere is made of two hollow copper hemispheres filled with 5 layers of rare-earth cobalt bricks. The copper wall thickness varies from 10 to 20 mm. The electrical conductivity of copper is about 4 times larger than that of sodium at 130$\deg$C. Once filled, the two hemispheres have been welded using an electron beam technique at the Laboratoire de l'Acc\'el\'erateur Lin\'eaire in Orsay. The assembly has finally been magnetized in a 5 T axial magnetic field, at the High Magnetic Field Laboratory in Grenoble. Outside the copper sphere, the magnetic field is found to be an axial dipole, aligned with the axis of rotation and pointing downwards, with:
$$
\vec{B}(r,\theta) = \frac{\mu_0 M}{4 \pi r^3}(2 \vec{e}_r \cos\theta + \vec{e}_{\theta} \sin\theta)
$$
in spherical coordinates, with $\theta$ the colatitude and $M=-700$ Am$^2$ the magnetic dipolar moment, yielding a magnetic field magnitude ranging from $0.41$ T at the poles of the inner sphere down to $0.008$ T at the equator of the outer sphere.

\begin{table}
\begin{center}
\begin{tabular}{cllc}
\hline
symbol & property & units  & value  \\
\hline
 $\rho$  & density & kg m$^{-3}$& $930$ \\
 $\sigma$& electrical conductivity& $\Omega^{-1}$ m$^{-1}$  &$9\;10^{6}$ \\
 $\nu$ & kinematic viscosity & m$^2$s$^{-1}$& $6.5 \;10^{-7}$ \\
 $\eta$ & magnetic diffusivity &m$^2$s$^{-1}$ & $0.087$\\
 $c$ & sound velocity & m s$^{-1}$ & $2600$ \\
\hline
\end{tabular}
\end{center}
\caption{Physical properties of liquid sodium at $130 \deg$C (from \cite{CEA}). The magnetic Prandtl number of sodium is $\Pm = \nu/\eta = 7.5 \; 10^{-6}$.}
    \label{tab:properties}
\end{table}

\subsection{Electric potentials and azimuthal velocity}

Electric potentials are measured at several latitudes on the rotating outer sphere (see Figure \ref{fig:sketch}). The electric potential is obtained by simply sticking copper electrodes into 1 mm-diameter 4 mm-deep blind holes drilled in the stainless steel wall. Stainless steel is a poor electric conductor as compared to liquid sodium and thus it does not affect the dynamically generated electric potential, but it is a good enough electric conductor for the impedance of electrode pairs to be much smaller than that of the acquisition system. The signals pass through slip-rings for acquisition in the laboratory frame, where they go through a simple RC anti-aliasing 215 Hz low-pass filter, before being sampled at 1000 sample/second with a PXI-6229 National Instruments acquisition card.

The electric potentials measured at the surface can be related to the azimuthal velocities beneath the Hartmann layer. Indeed, from the $\theta$-component of Ohm's law:

\begin{equation}
\vec{j} = \sigma (\vec{E} +\vec{U}\times\vec{B})
\label{eq:ohm}
\end{equation}
we deduce (assuming steady-state):

\begin{equation}
\frac{\Delta V}{a \Delta \theta}= U_{\varphi} B_{r} - \frac{j_{\theta}}{\sigma}
\label{eq:potj}
\end{equation}
where $\Delta V$ is the electric potential difference between two electrodes placed along the same meridian and separated by a latitudinal angle $\Delta \theta$, $U_{\varphi}$ is the azimuthal velocity of the fluid, $B_r$ the radial component of the magnetic field at the latitude of the electrodes, and $j_{\theta}$ the $\theta$-component of the electric current.
Further assuming that the latitudinal component of electric currents is negligible outside the Hartmann layer, and since the electric potential is constant across Hartmann layers \cite{mor90}, the potential difference measured at the surface can be directly related to the bulk velocity:

\begin{equation}
\Delta V = a \, \Delta \theta \, B_r \,U_{\varphi}
\label{eq:pot}
\end{equation}

Note that with our sign convention and since the magnetic dipole has its south pole in the northern hemisphere, we expect positive $\Delta V$ in that hemisphere when the inner sphere spins from West to East ($U_{\varphi} > 0$).

\subsection{Ultrasonic Doppler velocimetry and meridional circulation}

We measure radial profiles of the radial velocity of the fluid by ultrasonic Doppler velocimetry. At two different latitudes ($10\deg$ and $-20\deg$) on the rotating outer sphere, we install ultrasonic transducers in removable assemblies (see Figure \ref{fig:sketch}). The transducers operate at 4 MHz and their active part is 5 mm in diameter (type TR0405LS from Signal Processing, Switzerland). The slightly diverging ultrasonic beam enters liquid sodium after passing through a 1.4 mm-thick stainless steel wall, which is polished on the sodium side. 

The electric signals pass through the above-mentioned slip-ring system, to be analyzed in the laboratory frame. Note that the 36-track system is enclosed in an electromagnetic shield, to prevent contamination of the weak back-scattered echoes by ambient noise. The treatment of the ultrasonic pulses and echoes is performed on a DOP 2000 velocimeter (Signal Processing). The instrument issues a series of 8-cycles pulses and listens to echoes, which are back scattered by weak heterogeneities present in the fluid. The time elapsed between the pulse and the echo provides the distance of the scatterer, while the associated Doppler frequency shift yields the radial velocity at that distance. In this manner, radial profiles of the radial component of velocity can be obtained, and repeated every 30 ms or so.
Since liquid metals are opaque, ultrasound Doppler velocimetry is the most promising non-intrusive technique for mapping flows in MHD and dynamo experiments. However, the application to liquid metals raises specific issues and proves difficult: successful measurements in liquid gallium and sodium are recent \cite{bri01,eck02}.
While electric potentials at the surface of the sphere are sensitive only to the azimuthal velocities, we get constraints on the meridional circulation from the ultrasonic Doppler velocimetry.

\subsection{Dimensionless numbers}

Table \ref{tab:dimen} gives the typical values of the dimensionless numbers in our experiments. We note that the Joule dissipation time $\tau_J$ is very short. The Hartmann number $\Ha$ does not depend on the imposed rotation rates: it is large everywhere in the sphere, and the thickness of the regular Hartmann layers  $\delta_{\Ha}$ is less than 1mm. $U_a$ is the velocity of the Alfv\'en waves. When the outer sphere spins, the Ekman number is less than $2\;10^{-6}$ and the thickness of the regular Ekman layers less than 0.3 mm. The values of the Elsasser number $\Lambda$  and of $\lambda$   indicate that the magnetostrophic regime ($\Lambda \approx 1$ and $\lambda \ll 1 $ \cite{car02}) can be achieved in the interior of the sphere, while the flow is strongly influenced by the magnetic field near the inner sphere, and by global rotation near the outer sphere. Even for moderate differential rotation ($\Delta f = 1$ Hz), the magnetic Reynolds number $\Rm$ is larger than 1, meaning that the effective magnetic field depends on the flow itself. Note that the Rossby number $\Ro$ is simply defined as $\Ro= \Delta \Omega/\Omega  = \Delta f/f$. Finally, the values of the interaction parameter $\N$ indicate that the flow is strongly influenced by the Lorentz forces, except near the outer sphere, where $(\vec{u}\cdot\vec{\nabla})\vec{u}$ terms dominate. However, we note in the discussion section that even with a strong magnetic field, some inertial terms can play an important role.

\begin{table}
\begin{center}
\begin{tabular}{ccccccc}
\hline
symbol & expression & units  & \multicolumn{4}{c}{value} \\
\hline
 $a$  & outer radius & m& \multicolumn{4}{c}{$0.21$} \\
 $b$ & inner radius & m & \multicolumn{4}{c}{$0.07$} \\
 $\tau_J$ & $a^2/\pi^2\eta$ & s & \multicolumn{4}{c}{$0.05$} \\
 & & & \multicolumn{2}{c}{$\underline{\qquad B=B_i \qquad}$} & \multicolumn{2}{c}{$\underline{\qquad B=B_o \qquad}$}\\
 $\Ha$ & $aB/\sqrt{\mu_0\rho\nu\eta}$& &\multicolumn{2}{c}{5200} & \multicolumn{2}{c}{210}\\
 $U_a$ & $B/\sqrt{\mu_0\rho}$&m s$^{-1}$ & \multicolumn{2}{c}{5.8} & \multicolumn{2}{c}{0.2}\\
 $\delta_{\Ha}$ & $\sqrt{\mu_0\rho\nu\eta}/B$ & m &\multicolumn{2}{c}{$4\;10^{-5}$} & \multicolumn{2}{c}{$10^{-3}$}\\
  \hline
 & & & \multicolumn{2}{c}{$\underline{\qquad f = 1 \text{Hz}\qquad}$} & \multicolumn{2}{c}{$\underline{\qquad f = 5 \text{Hz} \qquad}$}\\

 $\E$ & $\nu/\Omega a^2$ & & \multicolumn{2}{c}{$2\;10^{-6}$}& \multicolumn{2}{c}{$5\;10^{-7}$}\\
 $\delta_{\E}$ & $\sqrt{\nu/\Omega}$ &m&\multicolumn{2}{c}{$3\;10^{-4}$}& \multicolumn{2}{c}{$1.4\;10^{-4}$}\\
 & & & $\underline{\: B=B_i \:}$ & $\underline{\: B=B_o \:}$&$\underline{\: B=B_i \:}$ & $\underline{\: B=B_o \:}$\\
 $\Lambda$ & $\sigma B^2/\rho \Omega$ & & 62 & 0.1& 12& 0.02\\
 $\lambda$ & $V_a/a \Omega$ & & 4.4 & 0.2& 0.9& 0.04\\
 \hline
 & & & \multicolumn{2}{c}{$\underline{\qquad \Delta f = 1 \text{Hz} \qquad}$} & \multicolumn{2}{c}{$\underline{\qquad \Delta f = 5 \text{Hz} \qquad}$}\\

 $U$ & $b\Delta \Omega$ &m s$^{-1}$&\multicolumn{2}{c}{0.44}& \multicolumn{2}{c}{2.2}\\
 $\Rm$ & $U a/ \eta$ &&\multicolumn{2}{c}{1.1}& \multicolumn{2}{c}{5.3}\\
 $\Re$ & $U a/ \nu$ &&\multicolumn{2}{c}{$1.4\;10^5$}& \multicolumn{2}{c}{$7\;10^5$}\\
 & & & $\underline{\: B=B_i \:}$ & $\underline{\: B=B_o \:}$&$\underline{\: B=B_i \:}$ & $\underline{\: B=B_o \:}$\\

 $\N$ & $\sigma a B^2/\rho U$ & & 180 & 0.3& 37& 0.06\\
 \hline
\end{tabular}
\end{center}
\caption{Typical values of the relevant parameters and dimensionless numbers for different imposed rotation frequencies $f = \Omega/2\pi$  and differential rotation frequencies  $\Delta f= \Delta\Omega/2\pi$. For the numbers that depend on the magnetic field strength, two values are given, the first one with $B=B_i=0.2$ T at the equator of the inner sphere, the second one with $B=B_o=0.008$ T at the equator of the outer sphere.}
    \label{tab:dimen}
\end{table}

\begin{figure}
  \centerline{\includegraphics[width=11cm]{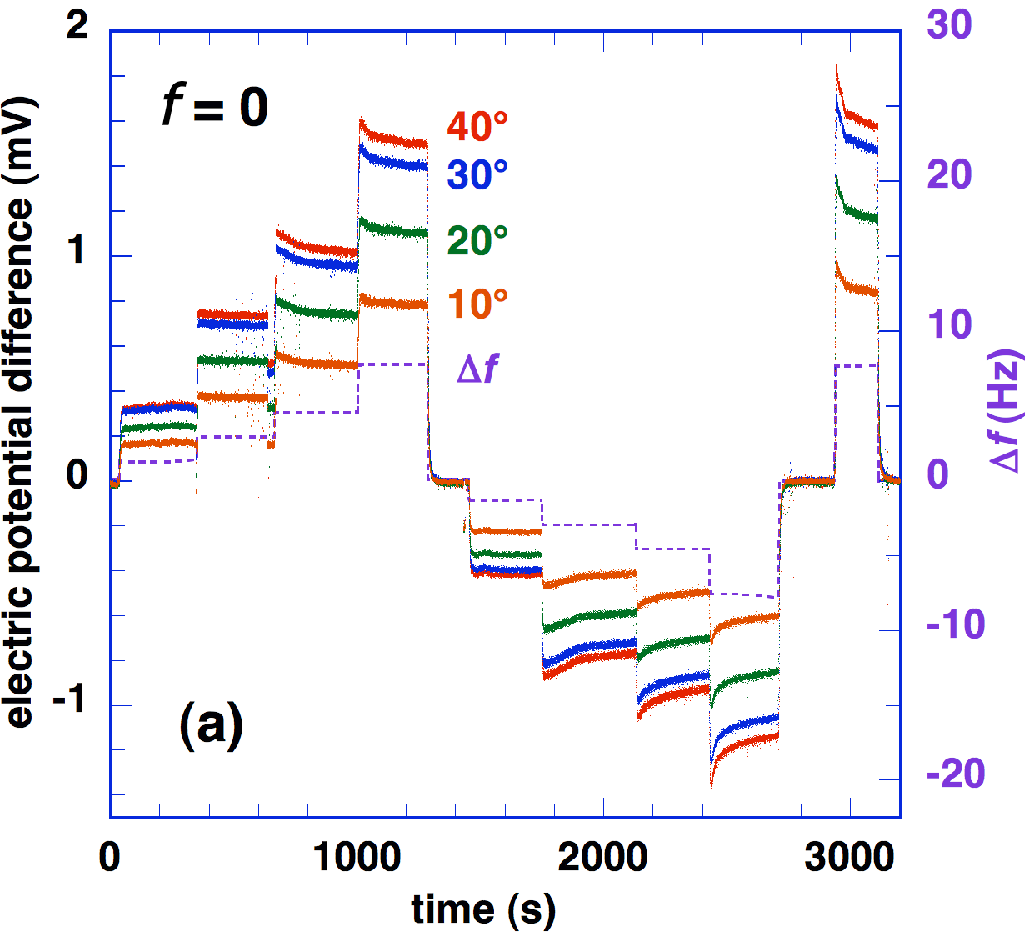}}
  \centerline{\includegraphics[width=12cm]{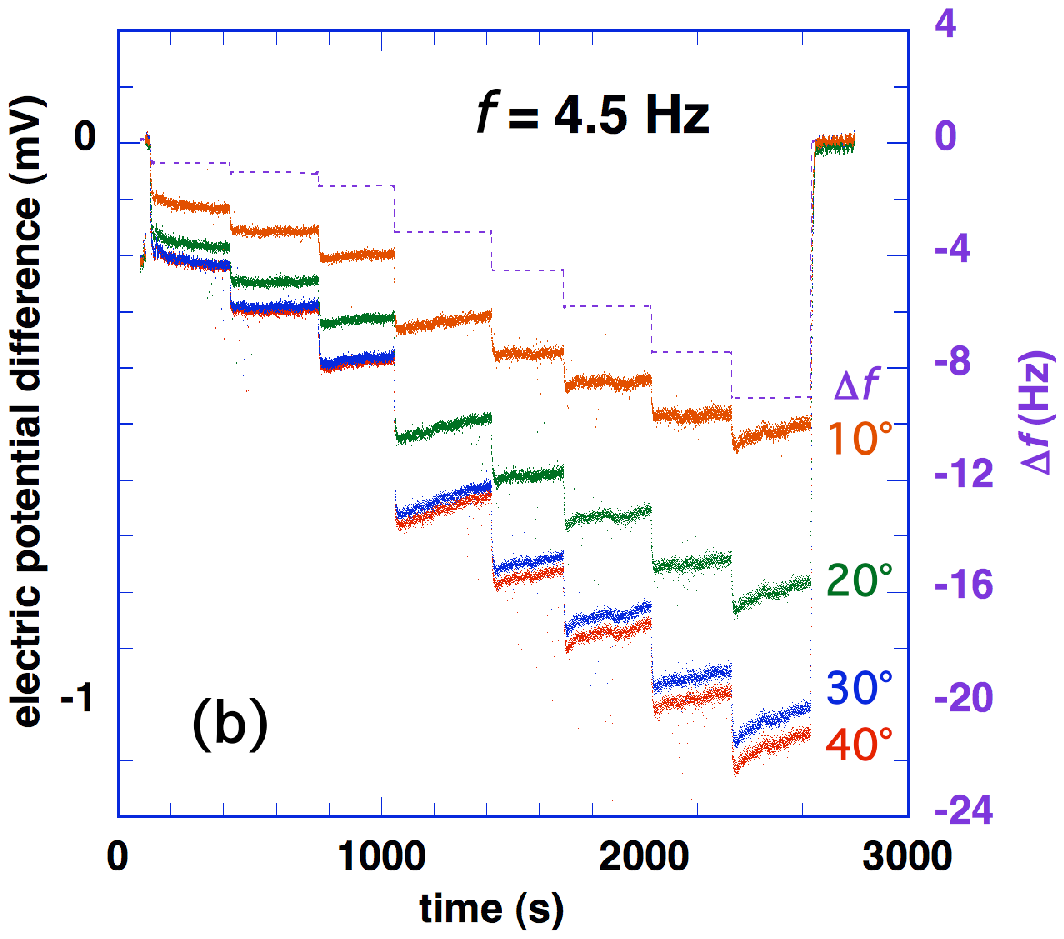}}
    \caption{Temporal records of the differences of electric potential between pairs of electrodes placed at different latitudes at the surface of the outer sphere. The rotation rate  $\Delta f$ of the inner sphere is raised in steps (dashed lines and right hand-side vertical axis). (a) outer sphere at rest ($f=0$). (b) rotating outer sphere ($f=4.5$ Hz).}
        \label{fig:pot}
\end{figure}

\section{Results}

\subsection{Azimuthal velocities from electric potentials at the surface of the outer sphere}

Figure \ref{fig:pot} shows temporal records of the differences of electric potential between several pairs of electrodes at the surface of the outer sphere. In Figure \ref{fig:pot}a, the outer sphere is at rest, and the rotation rate  $\Delta f$ of the inner sphere is raised in steps between -8 Hz and 8 Hz. During each step, the rotation rate is kept constant to within $5\; 10^{-3}$ during some 300 s. The electric signals are low-pass filtered to retain frequencies below 10 Hz. Each velocity step generates a jump in electric potential, forming a plateau sometimes preceded by a slow relaxation (probably of thermal origin). The signals in Figure \ref{fig:pot} demonstrate that the fluctuations of azimuthal velocity in this frequency range are very small compared to their mean value, and that the coupling between the inner sphere and the fluid is very strong (between steps, the fluid reaches its equilibrium velocity in less than 2 s). The amplitude of the signals is in the mV range.

In each pair, the electrodes are all 10$\deg$ apart, and the latitude of their mid-point ranges from $10\deg$ to $40\deg$. For $40\deg$, two pairs have been installed along two different meridians. Their signals are undistinguishable in Figure \ref{fig:pot}. The records show that the difference in electric potential increases with latitude, related to variations of the velocity field as the corresponding increase is known for the imposed radial magnetic field $B_r$ (see equation \ref{eq:pot}).

Figure \ref{fig:pot}b shows a similar temporal record, but this time the outer sphere spins at a constant rotation rate of 4.5 Hz during the whole sequence. The same behavior is observed.
For each imposed $\Delta f$ we retrieve the value of the electric potential plateau. 

We first concentrate on the signal measured at latitude $40\deg$.  We derive $U_{\varphi}$, using equation \ref{eq:pot} and divide it by $s\Delta\Omega$ , where $s$ is the cylindrical radius at that latitude. We can thus compare the azimuthal velocity of the fluid relative to that of solid body rotation (with the inner sphere) for different $f$ and  $\Delta f$.

\begin{figure}
  \centerline{\includegraphics[width=15cm]{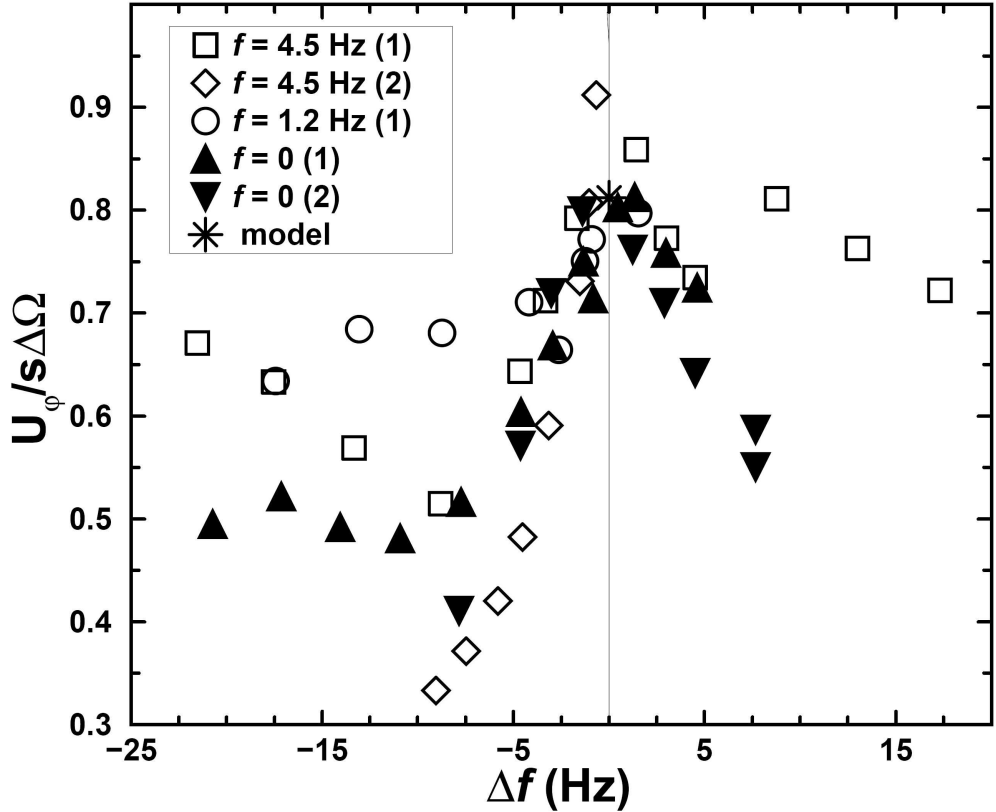}}
    \caption{Variation of the azimuthal velocity (normalized to that of solid body rotation with the inner sphere) at latitude $40\deg$, as a function of the rotation rate $\Delta f$ of the inner sphere, for various rotation rates $f$ of the outer sphere.  Two independent series (1) and (2) are presented. Typical uncertainties are below $10\%$ within a given series, but different series can be separated by a slightly larger amount. The star (model) at $\Delta f = 0$ is derived from the numerical model for $f=0$ as discussed in the text.}
        \label{fig:40}
\end{figure}
 
Figure \ref{fig:40} summarizes our results as a function of  $\Delta f$, for various $f$. We observe that, for a given $f$, the normalized velocity is maximum, reaching 0.8 to 0.9, when $\Delta f$ is minimum. We note that azimuthal velocity levels off at a higher value when global rotation is present, in particular for $\Ro>0$.

We now discuss the case without global rotation ($f=0$). The normalized azimuthal velocity decreases rapidly as $|\Delta f|$ increases, and seems to level off at a value of 0.5 for $|\Delta f|>10$ Hz. Our observations can be extrapolated to the $\Ro\rightarrow 0$ ($\Delta f\rightarrow0$) limit, for which numerical and analytical results are available.

The numerical simulations provide an excellent guide to interpret our results. We have run the numerical model described in \cite{dor02}, with parameters as close as possible to the experimental conditions: same Hartmann number (see Table \ref{tab:dimen}), electrical conductivities in the ratio 32/8/1 for inner sphere/fluid/outer shell, with appropriate thickness for the latter. The geometry of the applied magnetic field is the same as in our experiments (axial dipole). Figure \ref{fig:num} shows the contours of $U_{\varphi}/s\Delta\Omega$, meridional electric currents, and electric potential $V$ obtained numerically. At latitude $40\deg$, the Hartmann layer is well identified and is very thin. The actual azimuthal velocity beneath that layer at $40\deg$ is given for comparison with our experimental results in Figure \ref{fig:40} (`$*$' symbol).

\begin{figure}
  \centerline{\includegraphics[width=15cm]{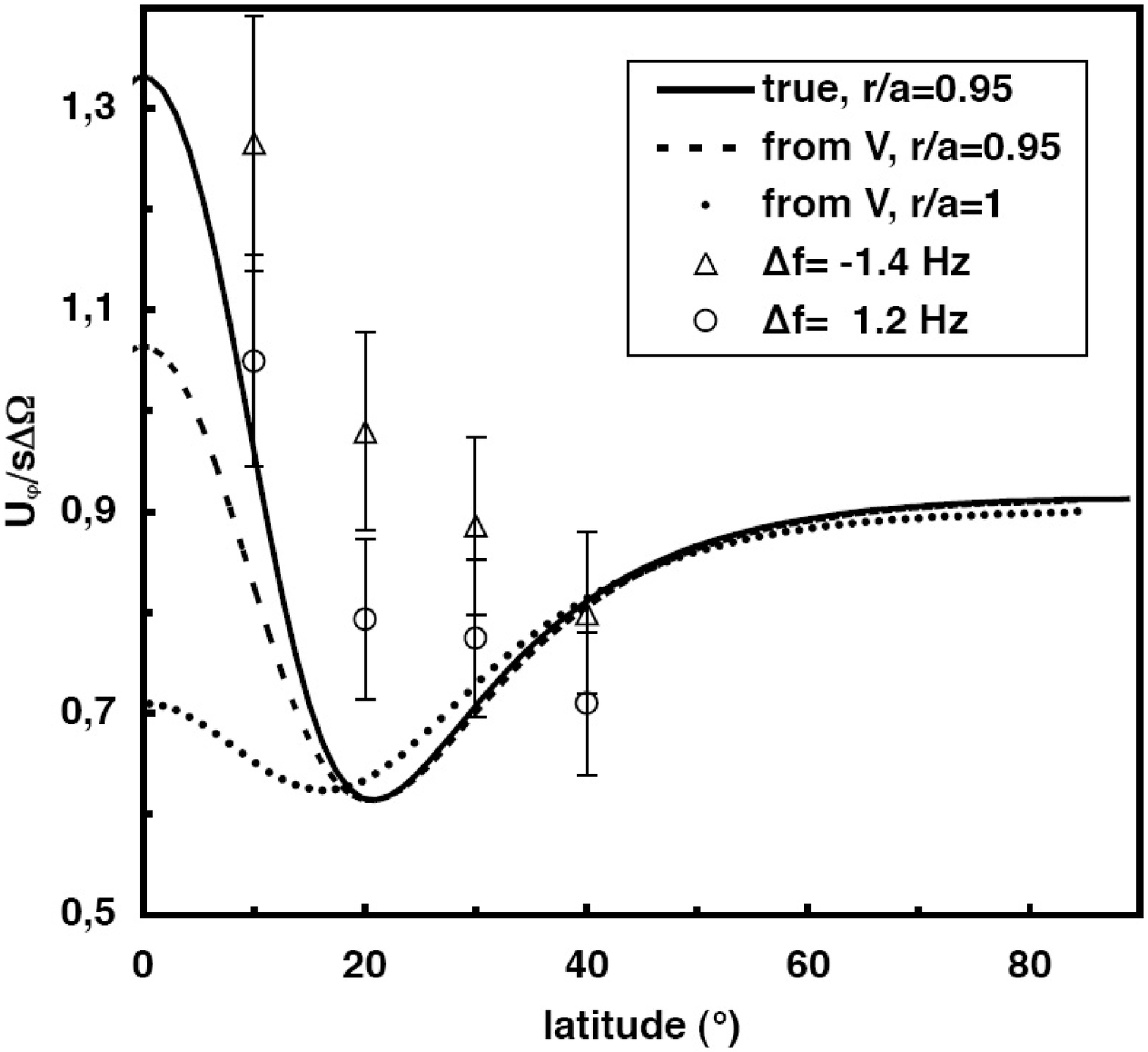}}
    \caption{Comparison between the azimuthal velocity beneath the Hartmann layer (solid line, $r/a=0.95$) and the azimuthal velocity inferred from the surface electric potential (dotted line, $r/a=1$) using equation \ref{eq:pot}, as a function of latitude, for the same numerical simulation as in Figure \ref{fig:num}. The dashed line is the velocity we would derive from the electric potential at $r/a=0.95$. The symbols with error bars are the experimental normalized angular velocities, inferred from the surface electric potential, for the lowest $|\Delta f|$ with $f=0$.}
        \label{fig:prof}
\end{figure}

\begin{figure}
  \centerline{\includegraphics[width=8cm]{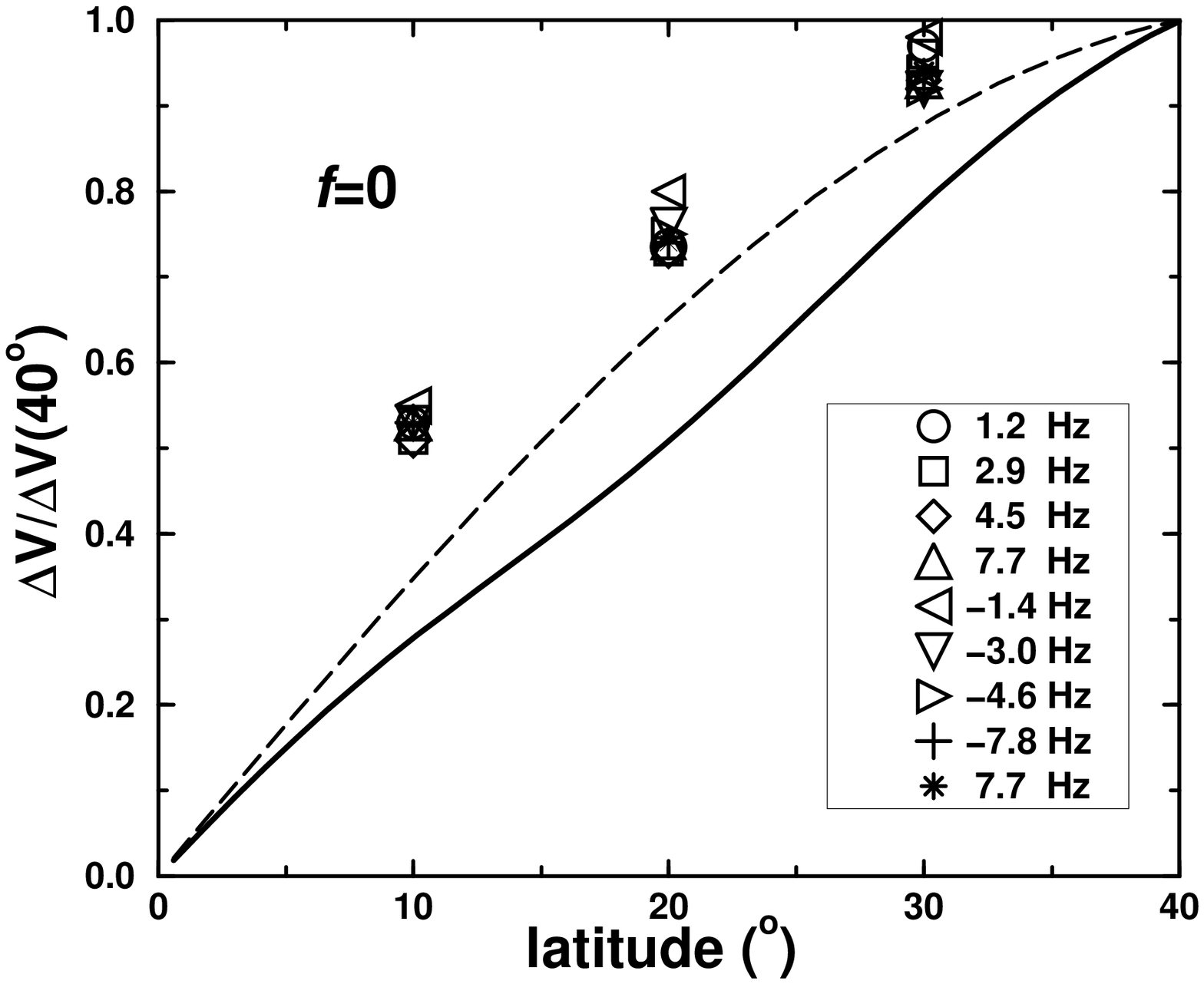} \includegraphics[width=8cm]{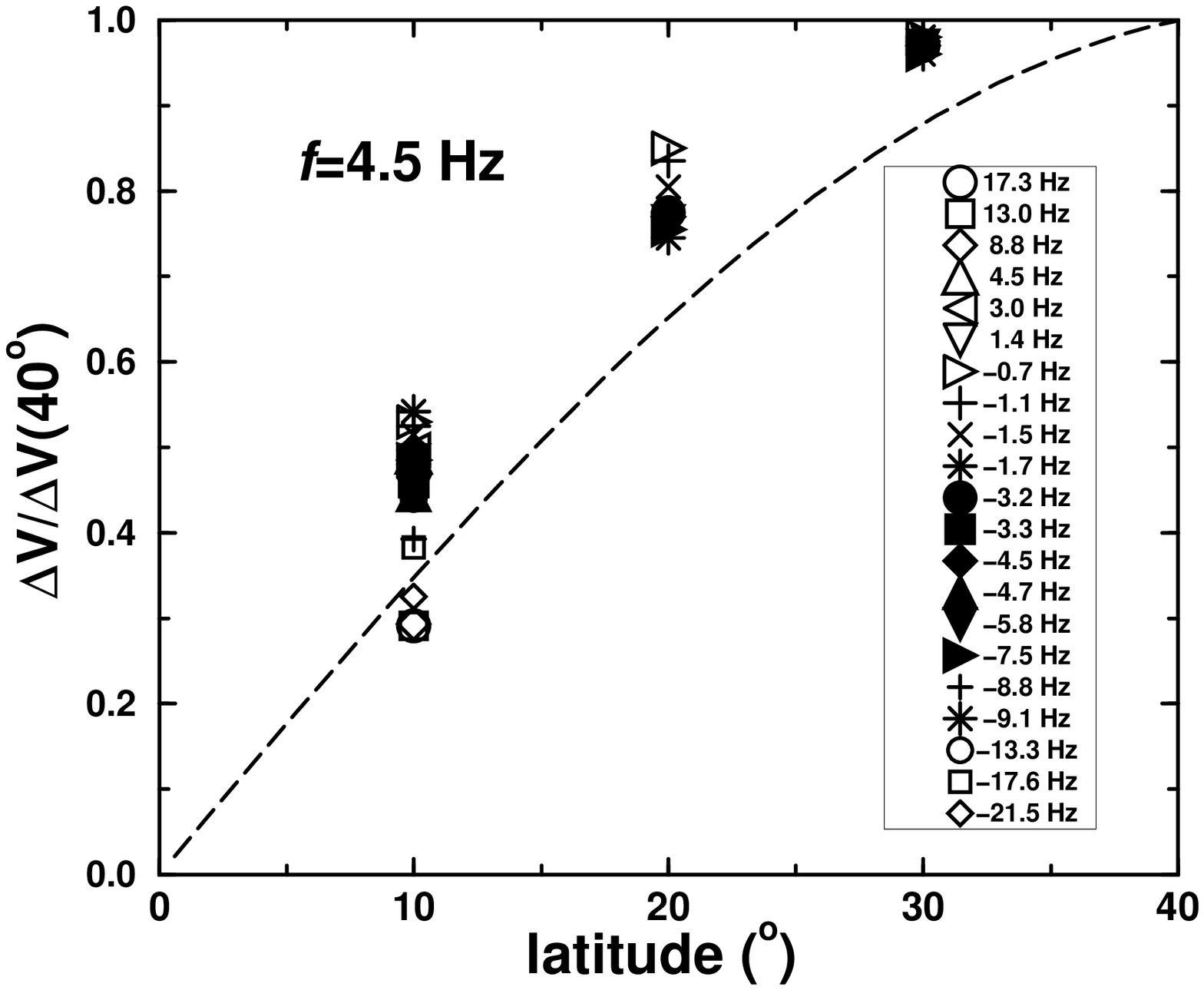}}
      \begin{picture}(0,0) (0,0)
    \put(30,173){(a)}
    \put(260,173){(b)}
  \end{picture}
    \caption{Latitudinal profiles of the differences of electric potential relative to that at latitude $40\deg$ for various rotation rates $\Delta f$ of the inner sphere. (a) outer sphere at rest ($f=0$). The solid line is the prediction of our numerical model. (b) rotating outer sphere ($f=4.5$ Hz). IThe dashed line indicates the potentials that would result from global solid body rotation with the angular velocity observed at latitude $40\deg$.}
        \label{fig:comp}
\end{figure}

We now investigate how the electric potential differences vary with latitude. The first move is to simply get $U_{\varphi}/s\Delta\Omega$ for the other latitudes using equation \ref{eq:pot}. This is what we have done in Figure \ref{fig:prof} for the experiments with the lowest $|\Delta f|$ and $f=0$. The data show that the angular velocity of the fluid increases towards the equator, reaching super-rotation angular velocities up to $30\%$ larger than solid body rotation with the inner sphere. However, the assumptions used to derive equation \ref{eq:pot} appear to break down when approaching the equator. This is illustrated by the curves derived from the above-mentionned numerical model. The true azimuthal velocity at $r/a=0.95$, outside the Hartmann layer, exhibits a slow decrease from the high latitudes down to $20\deg$, where it rises sharply to a maximum of more than $1.3$ at the equator. In contrast, the azimuthal velocity inferred from the electric potential at the surface using equation \ref{eq:pot} is almost twice smaller at the equator, while the differences are small above $40\deg$. There are two causes for this discrepancy and the clue is given by the third curve, which shows the azimuthal velocity that would be derived from equation \ref{eq:pot} using the electric potential at $r/a=0.95$. This time, the agreement with the true velocity is perfect down to latitude $20\deg$. Around the equator, the velocity is again underestimated. This demonstrates that the contribution of electric currents in the core (see equation \ref{eq:potj}), is indeed negligible at latitudes above $20\deg$, but that the electric potential is not quite constant across the Hartmann layer. In the Tropics, both approximations break down: the electric potential is not constant across the Hartmann layer, and the electric currents are not negligible beneath it. It is clear from Figure \ref{fig:num} that we are still under the influence of the $\Ha^{-1/2}$ thick layer at these latitudes.

Going back to our experimental measurements, we thus decide to directly plot the electric potential differences as a function of latitude. However, in order to focus on latitudinal variations, we normalize them with the potential difference measured at $40\deg$. Figure \ref{fig:comp}a shows our results for $f=0$ and $\Delta f$ varying from $-8$ to $8$ Hz. All profiles are remarkably similar despite a factor of almost 2 in their absolute level (see Figure \ref{fig:40}). The line derived from our numerical model falls well below our experimental points.

The measurements for $f=4.5$ Hz are given in Figure \ref{fig:comp}b. A similar behavior is observed.  Only those points with $|\Ro|= |\Delta f / f| > 2$ depart from the general trend, displaying smaller relative electric potential differences near the equator, below the line that corresponds to solid body rotation with the angular velocity observed at latitude $40\deg$, according to equation \ref{eq:pot}.

\subsection{Meridional circulation from ultrasonic Doppler velocimetry}

\begin{figure}
    \vspace{-2.5cm}
  \centerline{\includegraphics[clip=true, width=8cm]{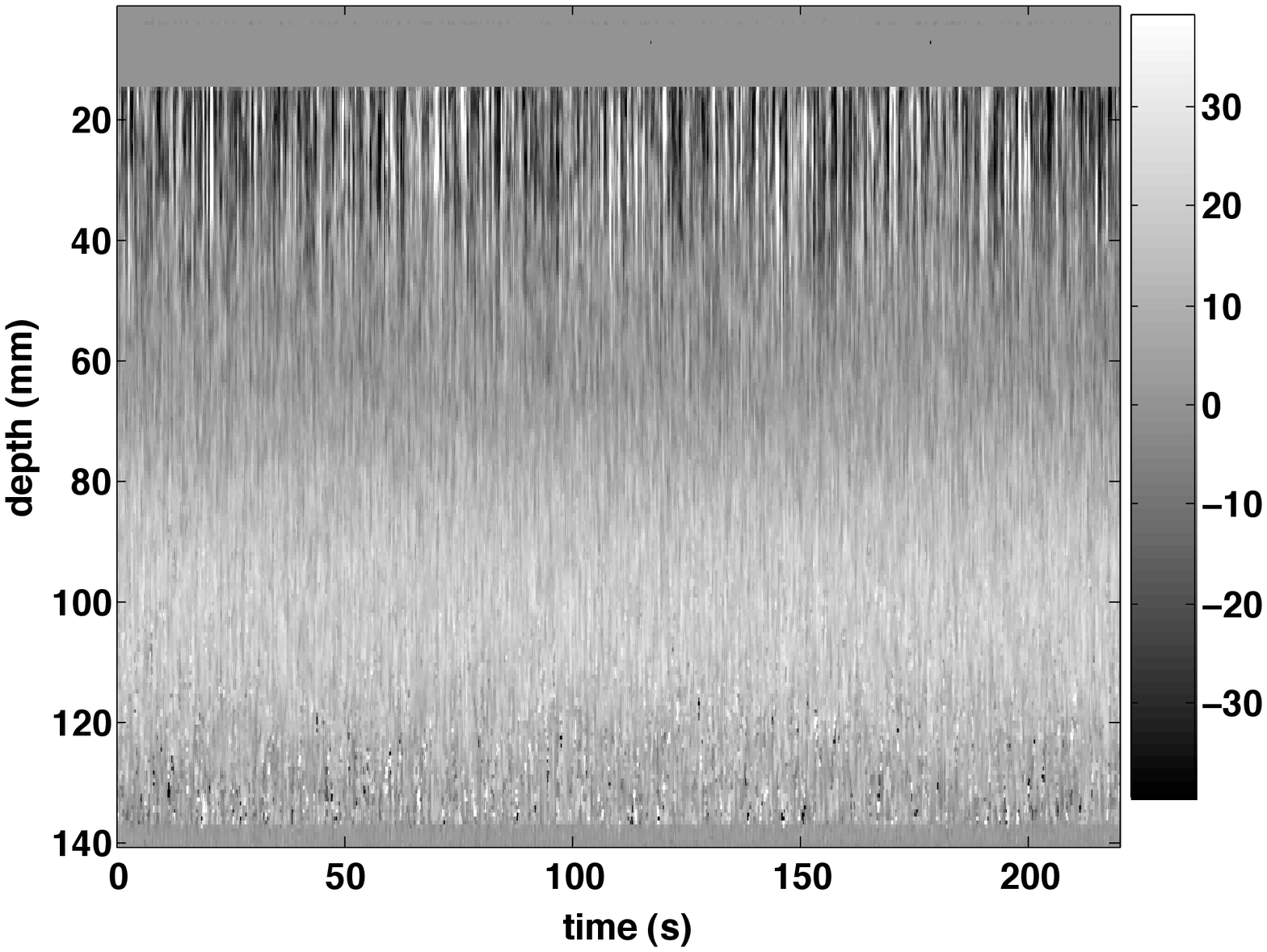} \includegraphics[clip=true, width=8cm]{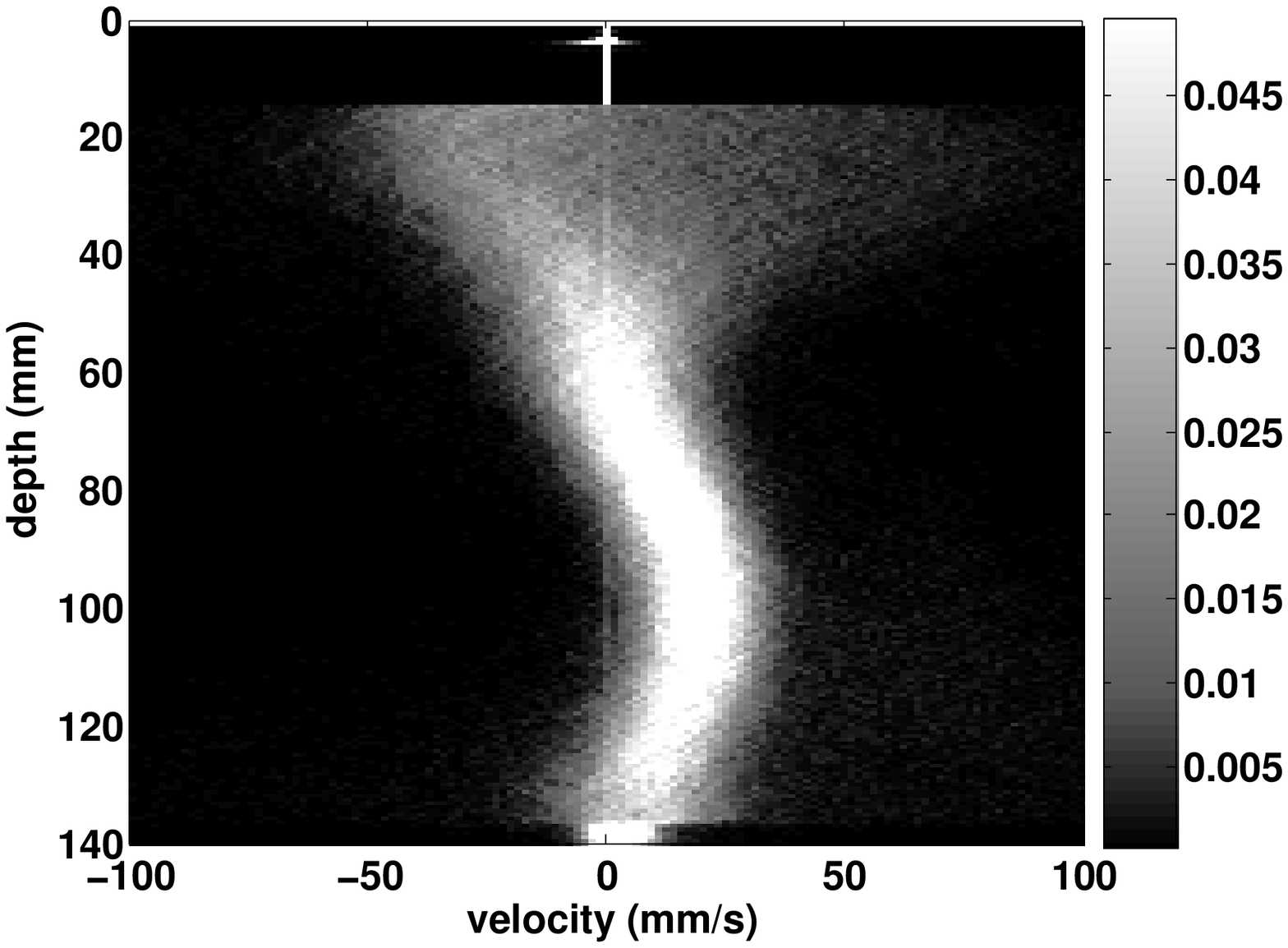}}
    \vspace{-5cm}
  \centerline{\includegraphics[clip=true, width=8cm]{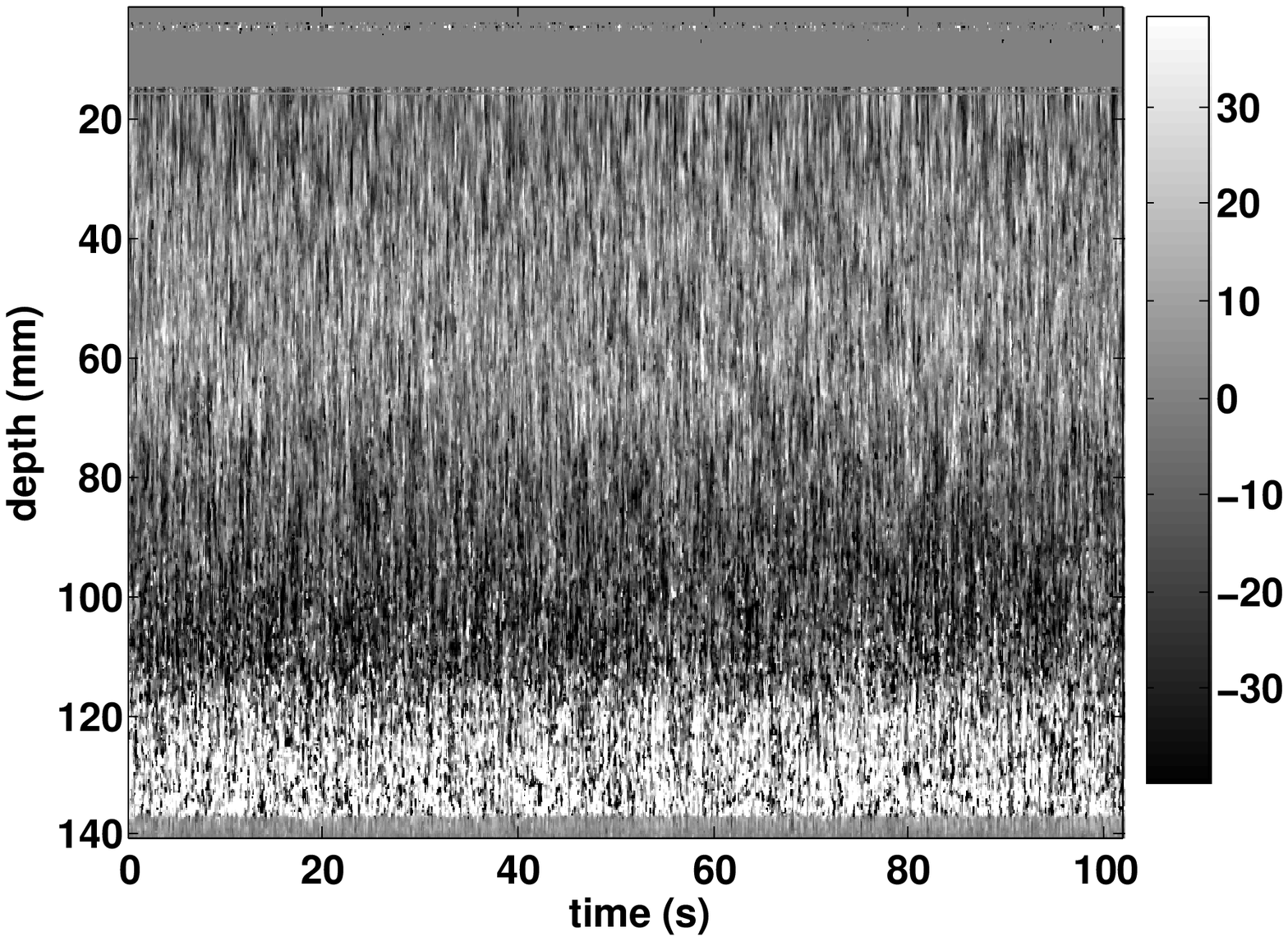} \includegraphics[clip=true, width=8cm]{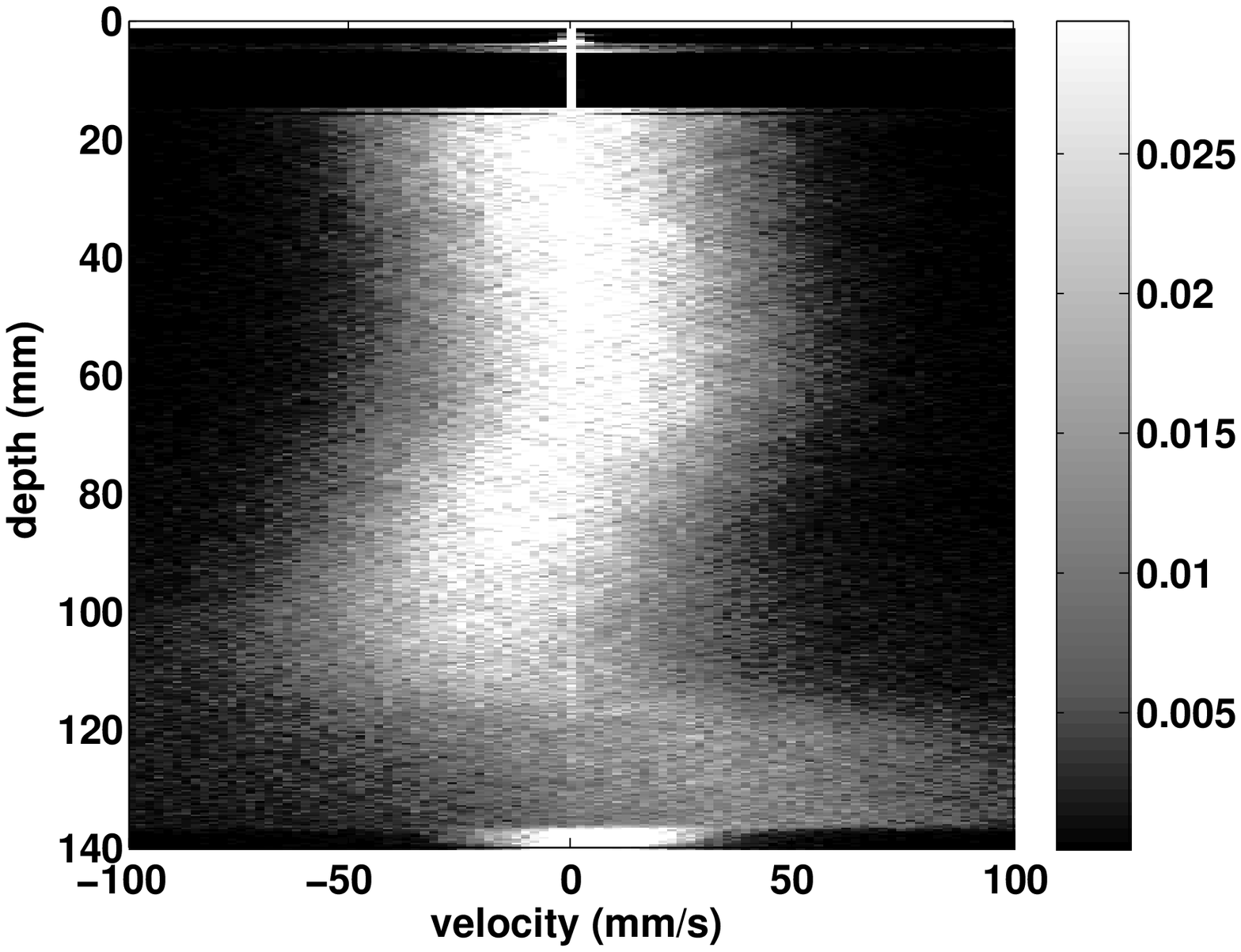}}
    \vspace{-2.5cm}
  \begin{picture}(0,0) (0,0)
    \put(20,305){(a)}
    \put(260,298){\textcolor{white}{(b)}}
    \put(20,145){(c)}
    \put(260,140){\textcolor{white}{(d)}}
  \end{picture}
    \caption{Ultrasonic Doppler velocimetry results: spatio-temporal plots and histograms of radial velocity as a function of depth.  The ultrasonic transducer is at latitude $-20\deg$ and at depth=$0$ at the top of the plot. The ultrasonic beam hits the inner sphere at a depth of 139 mm. On the left, the grey tones give the amplitude of radial velocity (scale in mm/s), centripetal in white and centrifugal in black. On the right, the grey tones are levels of the probability density function ($pdf$) of velocity. (a) and (b) outer sphere at rest and  $\Delta f=1.2$ Hz. Note the strong oscillations around depth=30 mm. (c) and (d) $f=4.7$ Hz and $\Delta f=3.0$ Hz.}
        \label{fig:dop}
\end{figure}

We now turn to the constraints brought on the meridional circulation by the analysis of radial profiles of radial velocity, obtained by ultrasonic Doppler velocimetry. Note that we cannot measure azimuthal velocities with this technique in the present set-up. Figure \ref{fig:dop} shows two examples of spatio-temporal plots and corresponding probability density functions ($pdf$) of velocity as a function of depth. Two different behaviors are observed. When the outer sphere is a rest (top row), strong and rapid oscillations extend from the outer surface down to about 40-50 mm. The corresponding $pdf$ is very asymmetric: centripetal velocities reaching much larger values than centrifugal ones. Smooth stable centripetal velocities are measured from 80 mm down to the inner sphere. The equatorial oscillations are reminiscent of the jet instabilities observed in spherical Couette flow in the absence of a magnetic field.
The rotation of the outer sphere kills the equatorial oscillations (bottom row), and the $pdf$ is Gaussian at all depths. 

\begin{figure}
  \centerline{\includegraphics[width=7.7cm]{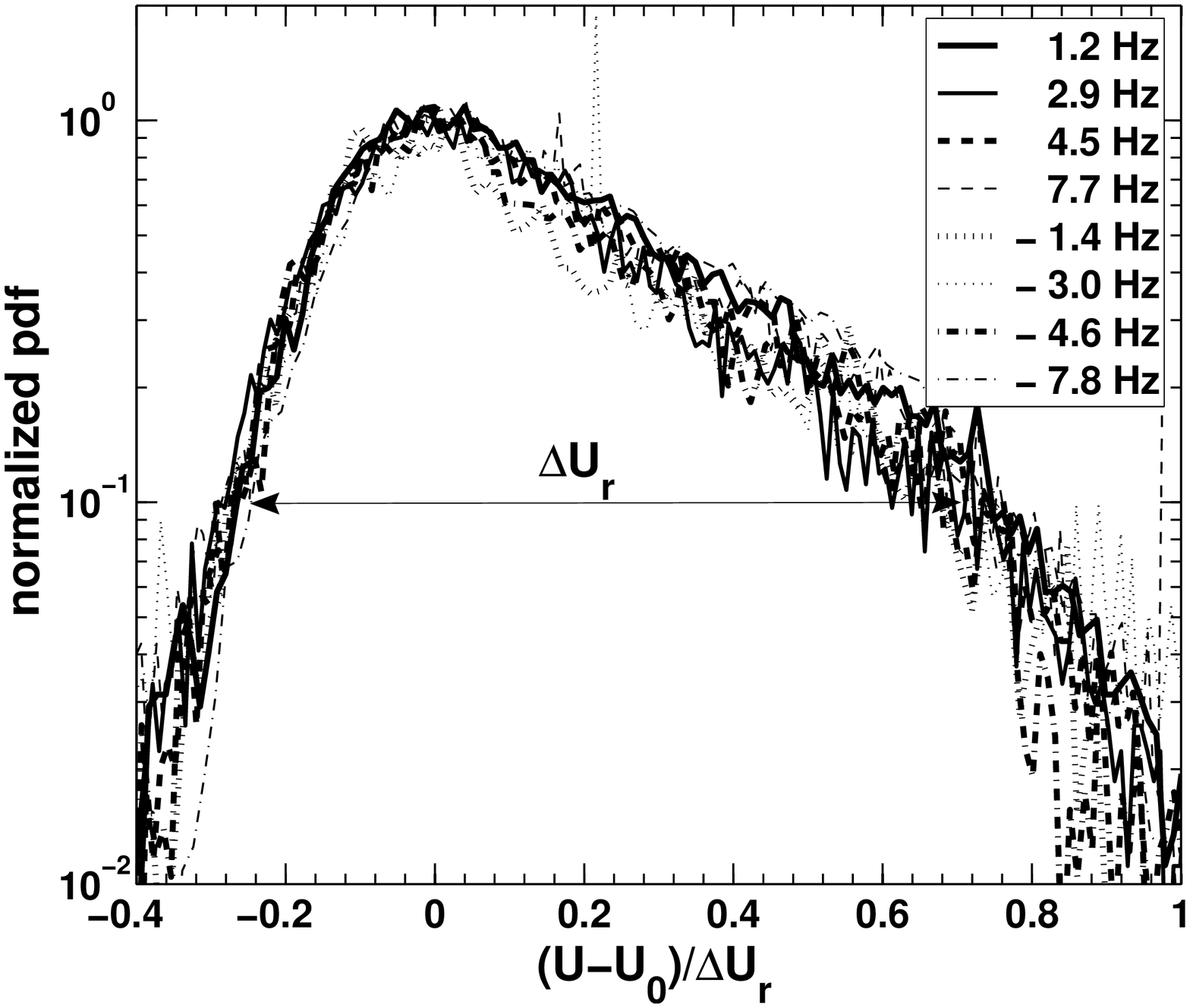} \includegraphics[width=8.3cm]{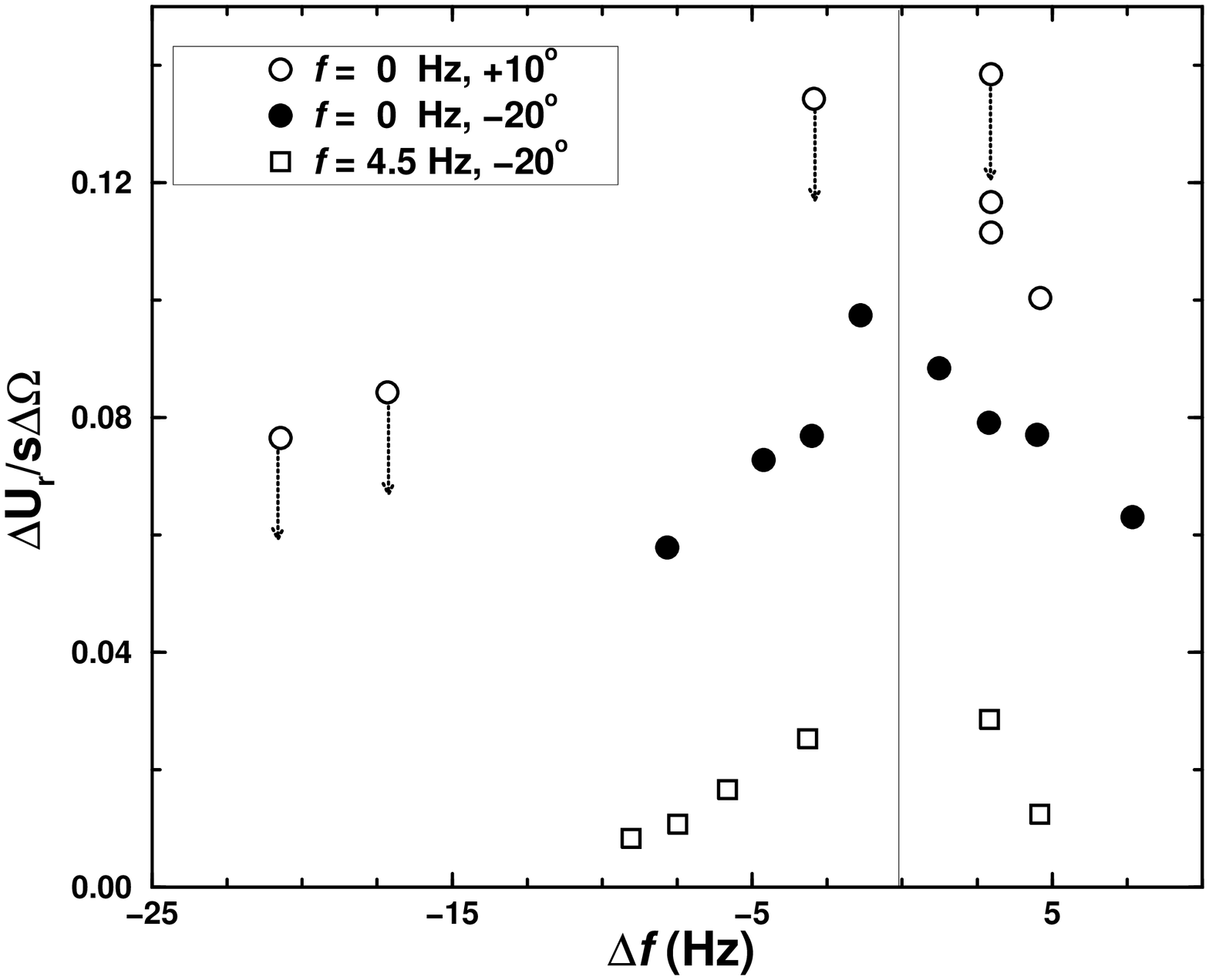}}
  \begin{picture}(0,0) (0,0)
    \put(30,40){(a)}
    \put(260,40){(b)}
  \end{picture}
    \caption{(a) normalized $pdf$ of radial velocities for $f=0$ at a depth of 27 mm, from ultrasonic Doppler velocimetry. Once rescaled with each width $\Delta U_r$ and shift $U_0$, all $pdf$ collapse on a single curve. (b) $\Delta U_r/s \Delta \Omega$ as a function of $\Delta f$. Circles: $f=0$. The full symbols are for a transducer at a latitude of $-20\deg$, while it is at $10\deg$ for the open symbols. The arrows indicate the shift to be applied to the data because different instrumental settings have been used for these data points. The squares are for $f=4.5$ Hz and show that the equatorial oscillations are suppressed by global rotation.}
        \label{fig:pdf}
\end{figure}

Getting back to the case with an outer sphere at rest, we find that the amplitude of the equatorial oscillations increases with $|\Delta f|$. Around the depth of 30 mm, we measure the spread  $\Delta U_r$ for which the $pdf$ has dropped down to one tenth of its maximum value. When the $pdf$ are rescaled using  $\Delta U_r$ and an offset $U_0$, they all have the same shape, as shown in Figure \ref{fig:pdf}a. We then plot  $\Delta U_r/s \Delta \Omega$ as a function of  $\Delta f$ (Figure \ref{fig:pdf}b) and observe the same kind of pattern as for azimuthal velocities: the ratio is maximum for small values of  $\Delta f$. The ultrasonic probe has been installed at two different latitudes: $+10\deg$ and $-20\deg$. When we compare the two series of results, we find that higher  $\Delta U_r/s\Omega$ values characterize latitude $+10\deg$. Since the equatorial oscillations probably result from instability of the equatorial azimuthal jet, this observation is in agreement with an increase of differential rotation towards the equator.

When the outer sphere rotates, the equatorial oscillations disappear, even when the Rossby number $\Ro= \Delta f/f$ gets down to -2.

\section{Discussion}

We have conducted experiments on the spherical Couette flow with global rotation and an imposed dipolar magnetic field. The working fluid is liquid sodium, and we have developed several instrumental techniques to characterize the flow. Surface electric potentials convey crucial informations about the azimuthal velocities in the fluid. At latitude $40\deg$, the dimensionless angular velocities $U_{\varphi}/s\Delta\Omega$ we retrieve reach 0.9. They are maximum for small differential rotation rates. When the outer sphere is at rest, the extrapolation to $\Delta f=0$ is in excellent agreement with the predictions of our numerical model. This numerical model exhibits super-rotation. It uses the same dipolar magnetic field and Hartmann number as in the experiments, and takes into account the finite electrical conductivities of the inner and outer walls. 

However, the latitudinal variation of electric potentials measured in the experiments differs markedly from the numerical predictions, and exhibit larger-than-expected values at low latitudes, which are not explained by the finite conductivity of the outer wall (correctly taken into account in the numerical model). 
Under the assumptions of the numerical model -- steadiness, equatorial symmetry, absence of non linearity for the magnetic field ($\Rm \ll 1$) and for the velocity field (Re $\ll 1$) -- these electric potential measurements indicate a much stronger super-rotation of the fluid than predicted numerically. 

The difference could be due to the meridional flow: while it is absent from the numerical solution for $f=0$, we measure radial velocities reaching about $10\%$ of the azimuthal velocities. The meridional flow advects the magnetic field lines, producing additional Lorentz forces. Thus, the meridional flow could be driven by centrifugal forces (one component of the Reynolds stresses) resisted by magnetic forces. According to this scenario, magnetic field lines and the associated region of super-rotation would be pushed closer to the surface in the equatorial region. The electric potentials measured at the surface could then directly reveal the super-rotation region.
The numerical results of Cupal et al. \cite{cup03} suggest that these additional Lorentz forces could also amplify the super-rotation.

Even when the outer sphere is at rest, we obtain slightly different results depending upon the direction of differential rotation. The equatorial symmetry could be broken by electric currents crossing the equatorial plan. This hypothesis could be tested by measuring the electric potential in both hemispheres in future experiments.

Finally, we note that the ultrasonic Doppler velocity measurements show that the flow is not steady in the equatorial region, except when the outer sphere spins. Instead, rapid oscillations of the radial velocity are observed, even for the lowest $|\Delta f|$.

The latitudinal profiles of angular velocity inferred from the electric potentials are remarkably similar across a wide range of parameters. This seems to indicate that a rather universal behavior is behind all this. More numerical modelling, relaxing the assumptions of \cite{dor98,dor02} and close to the experimental parameters are needed to explain our experimental findings.

We have obtained excellent results with ultrasonic Doppler velocimetry. Only radial velocities can be measured in the present set-up, but we are currently adapting it to accommodate measurements of the azimuthal velocities. This direct measurement will give us the amplitude and the radial distribution of the time-averaged angular velocity at a given latitude. Measurements of the induced toroidal magnetic field inside the sphere could also help characterize the latitudinal variation of the angular velocities.

Global rotation appears to enhance super-rotation, in particular at large $\Delta f$. As a consequence, very large azimuthal velocities are inferred from electric potentials, yielding $\Rm$ values in excess of $30$, with a power input below 4 kW. This result and the observation of a sizeable meridional circulation make the project of building a dynamo experiment based on the rotating spherical Couette flow appealing \cite{car02}.

\section{\bf Acknowledgments}
We are thankful to Dominique Grand and his colleagues from $SERAS$ who conducted the design study of the $DTS$ experiment. We are indebted to Antoine Alemany, Christian Chillet and Robert Stieglitz for their contributions to this design. We thank Alexandre Fournier for useful comments. The Laboratoire de l'Acc\'el\'erateur Lin\'eaire in Orsay and the High-Magnetic Field Laboratory in Grenoble kindly provided access to their equipments. The $DTS$ project is supported by Fonds National de la Science, Institut National des Sciences de l'Univers, Centre National de la Recherche Scientifique, R\'egion Rh\^ one-Alpes and Universit\'e Joseph Fourier.

\end{document}